\documentstyle[aps,prb,multicol,epsfig]{revtex}

\begin{document}
\title{Stability and electronic structure of the complex
K$_2$PtCl$_6$ structure-type hydrides}
\author{S.V. Halilov and D.J. Singh}
\address{
Center for Computational Materials Science, \\
Naval Research Laboratory, Washington, DC 20375-5000, USA}
\author{M. Gupta}
\address{
Thermodynamique et Physico-Chimie d'Hydrures et Oxydes, EA3547, Batiment 415, \\
 Science des Materiaux,
Universite Paris-Sud, 91405 Orsay, France}
\author{R. Gupta}
\address{Service de Recherches de Metallurgie Physique,
Commissariat a l'Energie Atomique, Centre d'Etudes de Saclay, \\
91191 Gif Sur Yvette Cedex, France}
\date{\today }
\maketitle

\begin{abstract}
The stability and bonding
of the ternary complex K$_2$PtCl$_6$ structure hydrides
is discussed using first principles density functional calculations.
The cohesion is dominated by ionic contributions, but ligand field
effects are important, and are responsible for the 18-electron
rule. Similarities to oxides are discussed in terms of the electronic
structure. However, phonon calculations for Sr$_2$RuH$_6$ also
show differences, particularly in the polarizability of the RuH$_6$
octahedra. Nevertheless, the yet to be made compounds
Pb$_2$RuH$_6$ and Be$_2$FeH$_6$ are possible ferroelectrics.
The electronic structure and magnetic properties of the decomposition product,
FeBe$_2$ are reported.
Implications of the results for H storage are discussed.
\end{abstract}


\begin{multicols}{2}

\section{Introduction}

The complex hydrides, $D$$M$H$_6$, $D$=Mg,Ca,Sr,Eu and
$M$=Fe,Ru,Os form in the cubic ($Fm3m$,\#225), K$_2$PtCl$_6$ structure.
\cite{moyer,didisheim,kritikos,huang,yvon,moyer2}
This structure has $D$ on site 8$c$ (1/4,1/4,1/4), $M$ on 4$a$ (0,0,0),
and H on site 24$e$ ($x_{\rm H}$,0,0).
These compounds are of fundamental interest because of the unusual
structural motif and 
the interest in understanding resulting electronic structure,
and the bonding associated with it.
Furthermore, they could be of practical interest as potential
hydrogen storage materials. Mg$_2$FeH$_6$ has very
high volume and mass storage efficiency (150 g/l and 5.4 wt. \%),
but is too stable for reversible room temperature applications.
\cite{yvon,konst,bogdanovic}
In this regard,
understanding of the electronic structure and cohesion
may be helpful in finding modifications that improve the thermodynamics,
to produce a material for hydrogen storage in mobile applications.

The crystal structure may be regarded
as a cubic double perovskite $A_2 B B'$H$_6$, with $A=D$, $B$=$M$ and
$B'$=vacancy. Therefore, from a structural point of view, the compounds
consist of $M$H$_6$ octahedra, well separated by
presumably inert $D$ ions, whose role is to fill space and donate
charge to the $M$H$_6$ units. The cubic $Fm3m$ structure
is maintained for all these
compounds in spite of large variations of the $A$ and $B$ site cation radii,
in contrast to the structural distortions often found in oxide perovskites
and double perovskites.

These compounds are insulators, and,
like many of the complex hydrides, \cite{yvon}
follow the 18 electron rule, which says that the number of non-bonded
metal electrons plus the number of electrons in the metal-ligand
bonds should be 18. In the simplest view, this would correspond to
full $s$, $p$ and $d$ shells associated with the [$M$H$_6$]$^{4-}$
structural units. However, Miller and co-workers, \cite{miller}
have emphasized
the importance of ligand field effects in these complex hydrides, 
and, in fact, the calculated electronic structures for these materials
show band gaps within the $d$ manifolds, indicating
a more complex situation. \cite{orgaz}
In particular, insulating band structures, in qualitative accord
with experiment, resulting from
band gaps between crystal field split transition
metal $d$ manifolds were found in non-self-consistent
warped muffin-tin calculations, based on
the X$\alpha$ method with $\alpha=1$.

Here, we re-examine the electronic structures, which we
obtain using a full-potential,
self-consistent linearized augmented planewave (LAPW) method, and
use these results, along with
calculations of the formation enthalpies,
to discuss the bonding of these materials and
possible directions for modifying them to alter their stability.
In addition we discuss the hypothetical materials, Pb$_2$RuH$_6$,
which was studied as a potential ferroelectric, and Be$_2$FeH$_6$,
which is both in relation to ferroelectricity and to better understand their
stability.

\section{Approach}

As mentioned,
the calculations were done within the local density approximation (LDA)
using the general potential
linearized augmented planewave (LAPW) method. \cite{singh-book}
Local orbital extensions \cite{singh-lo}
were used to relax the linearization
errors for the transition metal $d$ states, and to treat the upper
semicore levels of the alkaline earth and transition metal atoms.
An LAPW sphere radius $r_{\rm H}$ = 1.1 $a_0$
was used for H in all the compounds.
Metal radii of 1.8 $a_0$ were used for Mg$_2$FeH$_6$ and
Ca$_2$FeH$_6$. For Sr$_2$FeH$_6$, metal radii, $r_{\rm Sr}$ = 2.0 $a_0$
and $r_{\rm Fe}$ = 1.8 $a_0$ were used for Sr and Fe, respectively.
For Ca$_2$RuH$_6$ and Sr$_2$RuH$_6$, metal radii of 1.95 $a_0$ were used.
For Mg$_2$RuH$_6$ we used $r_{\rm Mg}$ = 1.8 $a_0$ and
$r_{\rm Ru}$ = 1.9 $a_0$.
Well converged basis sets, defined by $r_{\rm H} G_{max}$=6.0, where
$G_{max}$ is the planewave cut-off were employed. The Brillouin zone
sampling during the iteration to self-consistency
was done using the special-{\bf k} points method with a 
$4\times 4\times 4$ mesh, which corresponds to 10 {\bf k}-points
in the irreducible wedge. Densities of states were generated using
a tetrahedron mesh of 145 {\bf k} points in the wedge.
Convergence was tested, both for the zone sampling and basis set size,
by repeating some calculations with higher $r_{\rm H} G_{max}$=7 and more 
{\bf k} points ($8\times 8\times 8$). Based on these tests, the present
convergence with respect to these parameters is better than
2 mRy/cell for total energies, and better than 1 mRy for band energies.

\section{Electronic Structure and the Eighteen Electron Rule}

Crystal field effects, which are important here, may be expected to be
sensitive to the H positions. Here we use LDA relaxation to determine
the H positions (given in Table \ref{table1}). As may be seen, they
are in good agreement with experiment for those compounds for which
neutron refinements are available. Table \ref{table1} also gives the
full symmetry $A_{g}$ Raman phonon frequency associated with the H internal
structural parameter. Raman and inelastic neutron scattering measurements
\cite{parker}
for Mg$_2$FeH$_6$ yield an experimental frequency of 1873 cm$^{-1}$,
in good agreement with the present LDA value of 1923 cm$^{-1}$.
The LDA frequencies
follow the reported trend for H bond stretching modes in infrared
data, \cite{yvon}
decreasing with increasing $D$ ionic radius, and increasing with
increasing $M$ atomic number. It is interesting to note the increase
in frequency from Sr$_2$RuH$_6$ to Sr$_2$OsH$_6$.
Generally, the ionic properties of 4$d$ and 5$d$ elements are very
similar due to relativistic contraction. The main difference, which
is also due to relativistic contraction, is that the $s$ states
are lower relative to the $d$ states in the 5$d$ series. This
leads to a higher position of the $d$
states relative to ligand states in compounds
where the transition elements are cations. The result is reduced
covalency. Since covalency softens ionic interactions, the result
is a stiffer lattice in 5$d$ ionic compounds relative to the corresponding
4$d$ compounds. A good example is the comparison of KNbO$_3$, which is
a good ferroelectric with KTaO$_3$, which has practically the same
lattice parameter, but is not ferroelectric.
\cite{singh-kt}

\begin{figure}[tbp]
\centerline{\epsfig{file=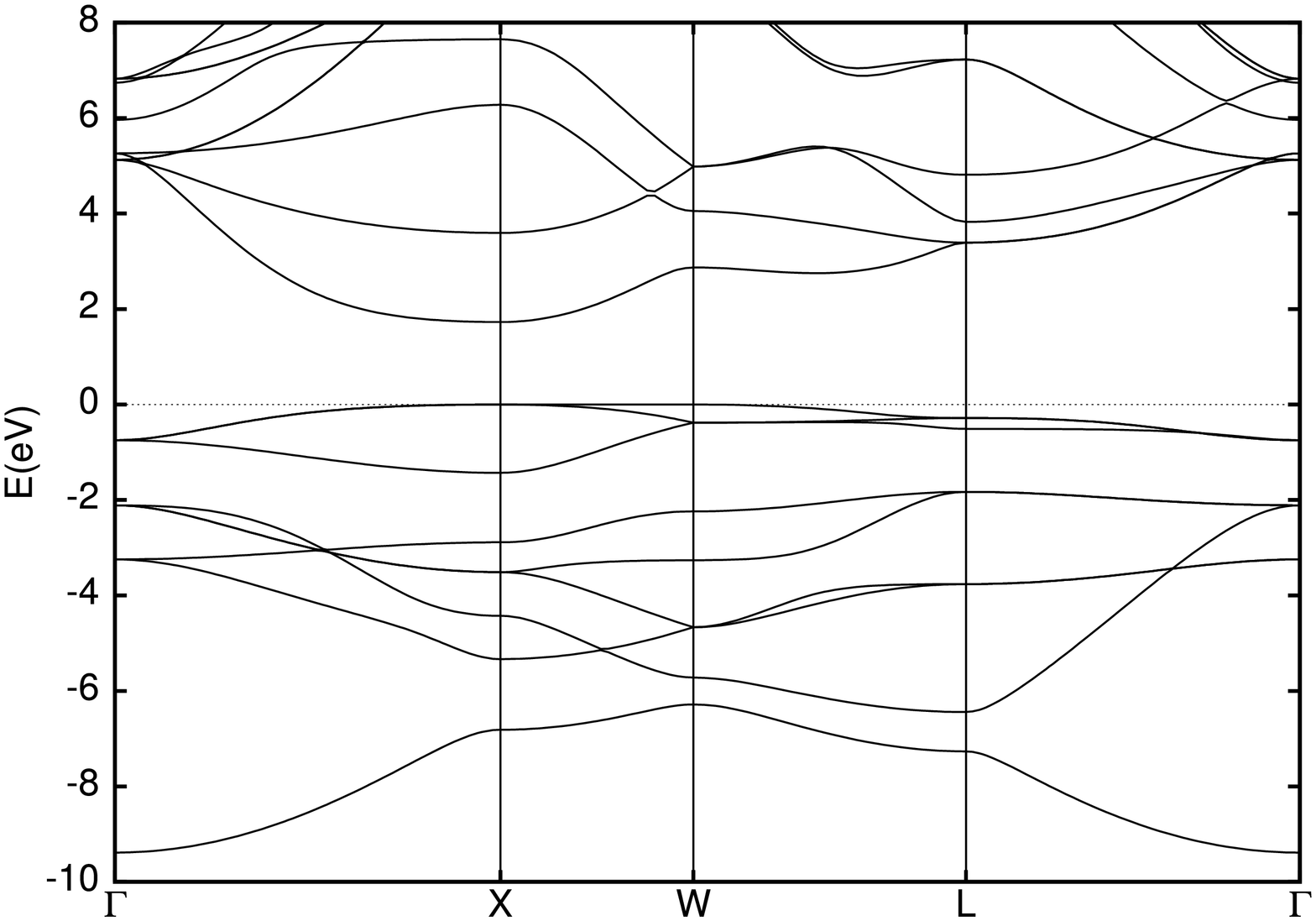,width=0.75\linewidth,angle=270,clip=}}
\vspace{0.1cm}
\centerline{\epsfig{file=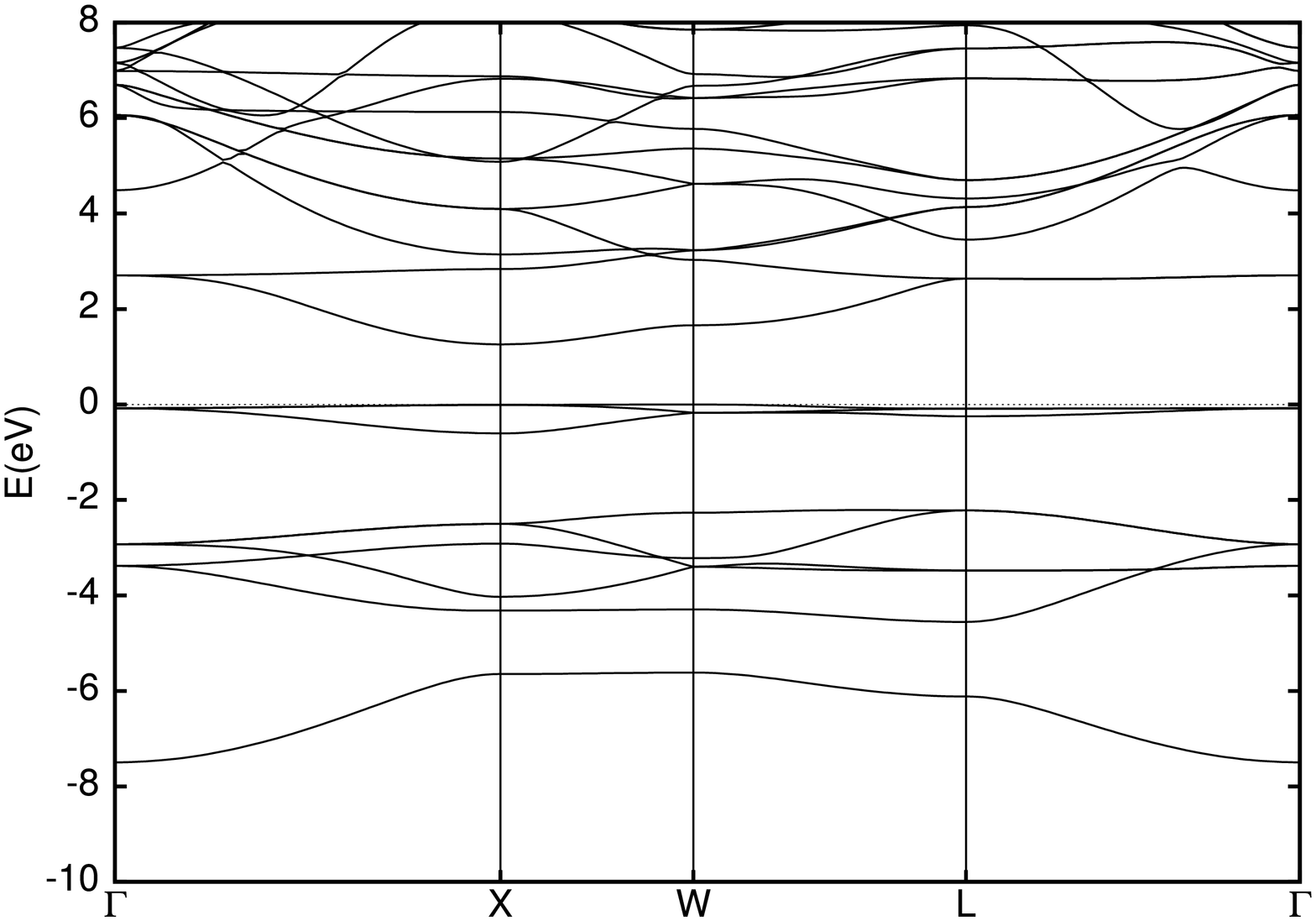,width=0.75\linewidth,angle=270,clip=}}
\vspace{0.1cm}
\centerline{\epsfig{file=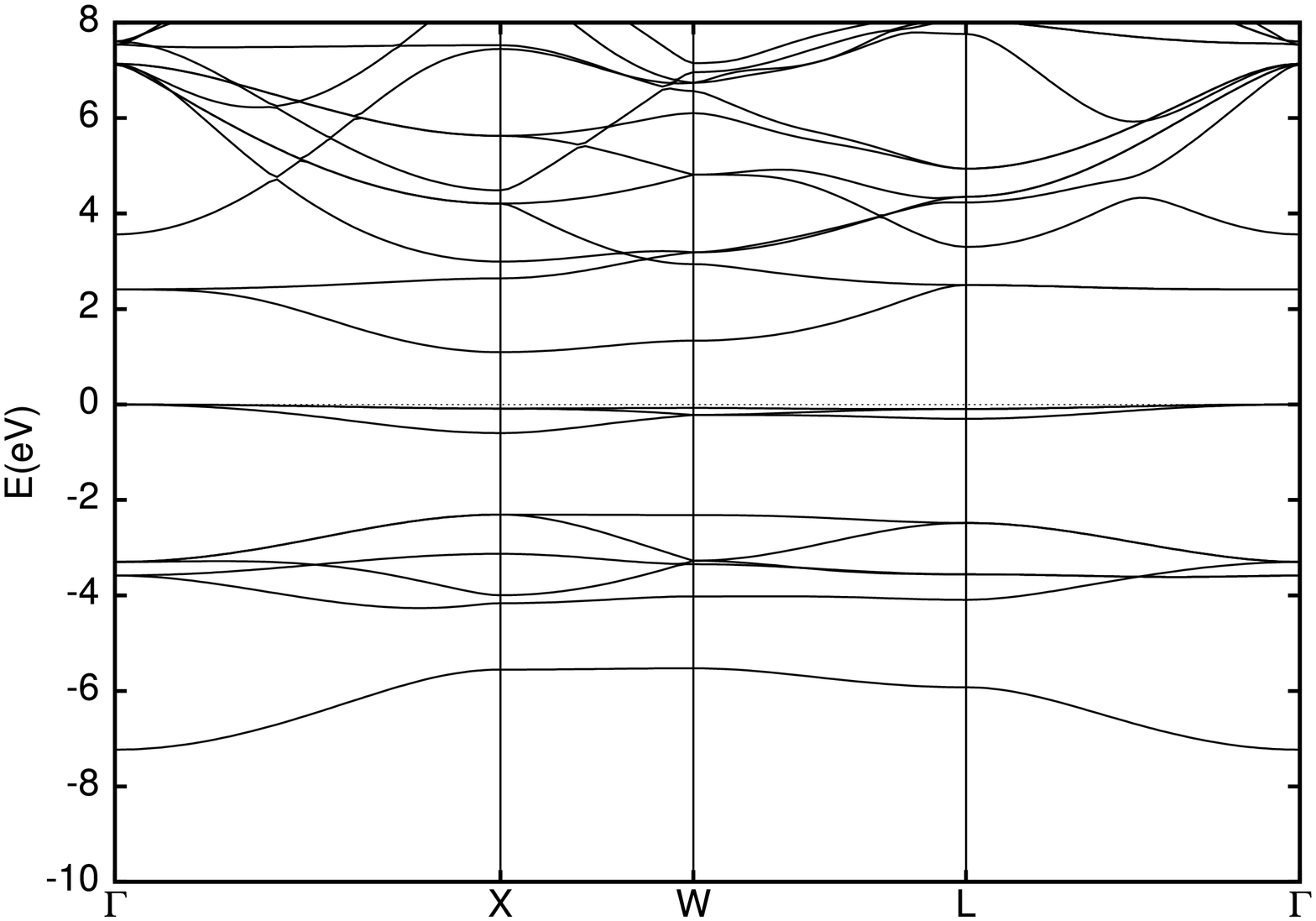,width=0.75\linewidth,angle=270,clip=}}
\vspace{0.1cm}
\caption{Band structure of Mg$_2$FeH$_6$ (top), Ca$_2$FeH$_6$ (middle)
and Sr$_2$FeH$_6$ (bottom), with the relaxed LDA H positions.
}
\label{bands-fe}
\end{figure}

\begin{figure}[tbp]
\centerline{\epsfig{file=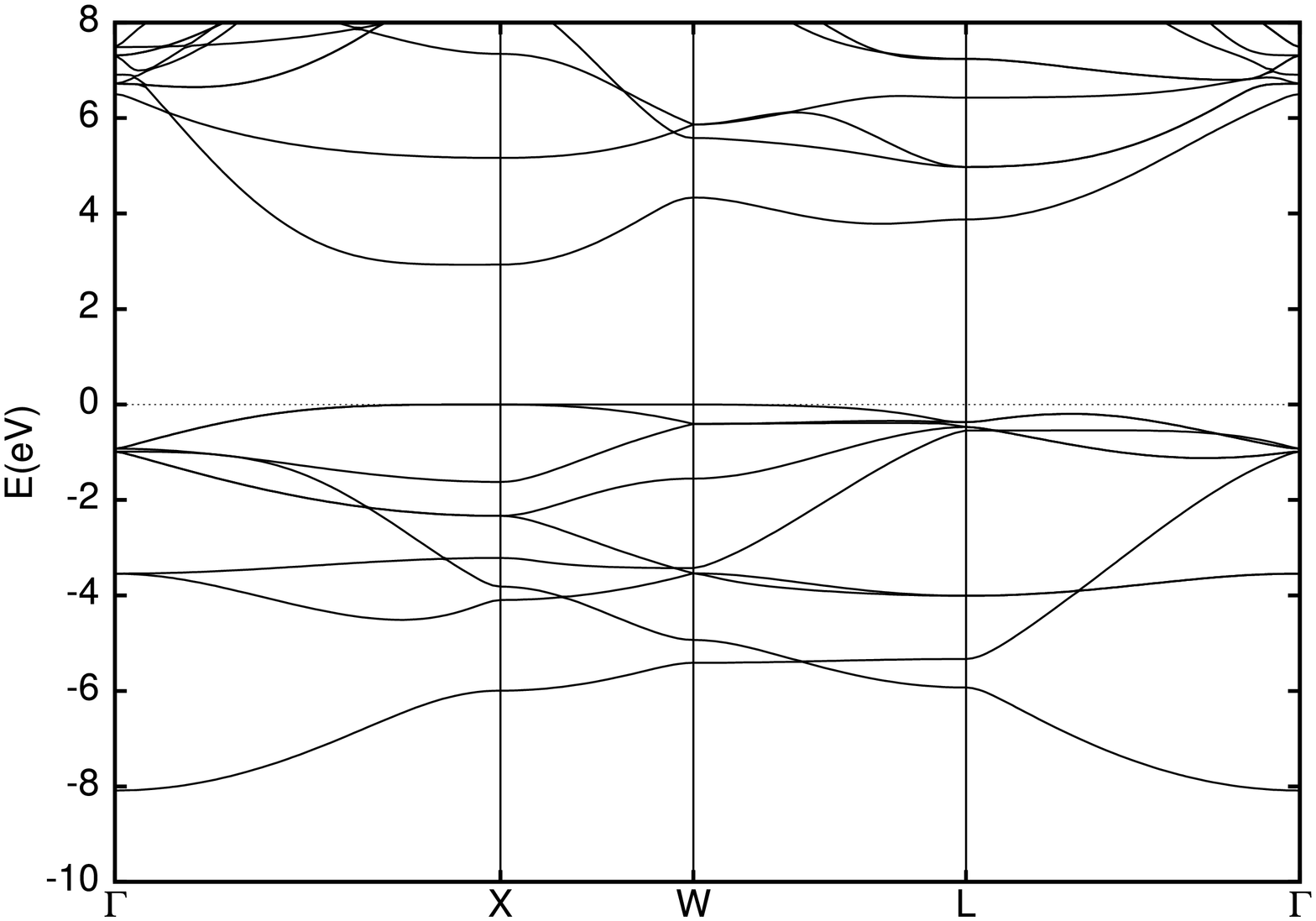,width=0.75\linewidth,angle=270,clip=}}
\vspace{0.1cm}
\centerline{\epsfig{file=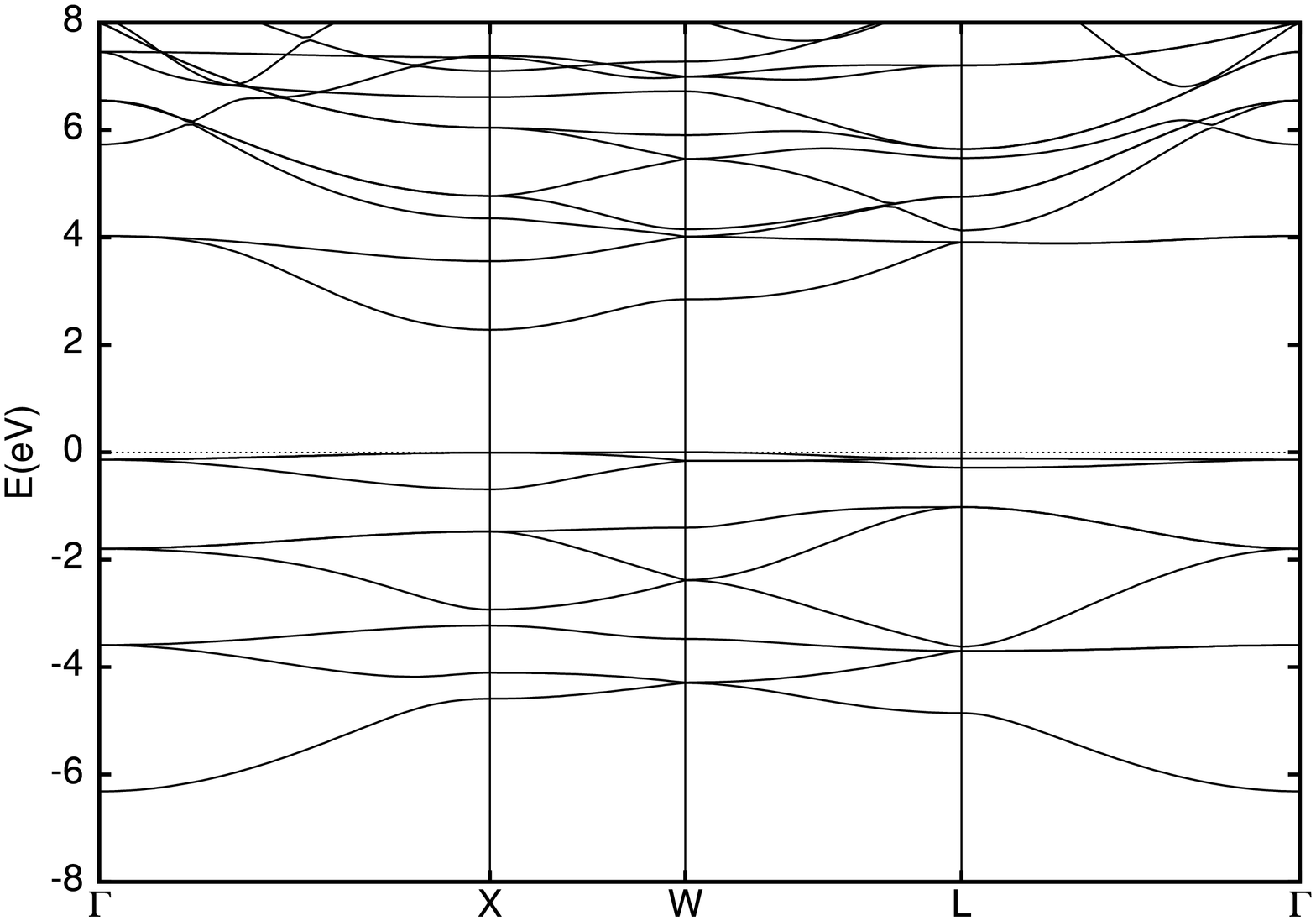,width=0.75\linewidth,angle=270,clip=}}
\vspace{0.1cm}
\centerline{\epsfig{file=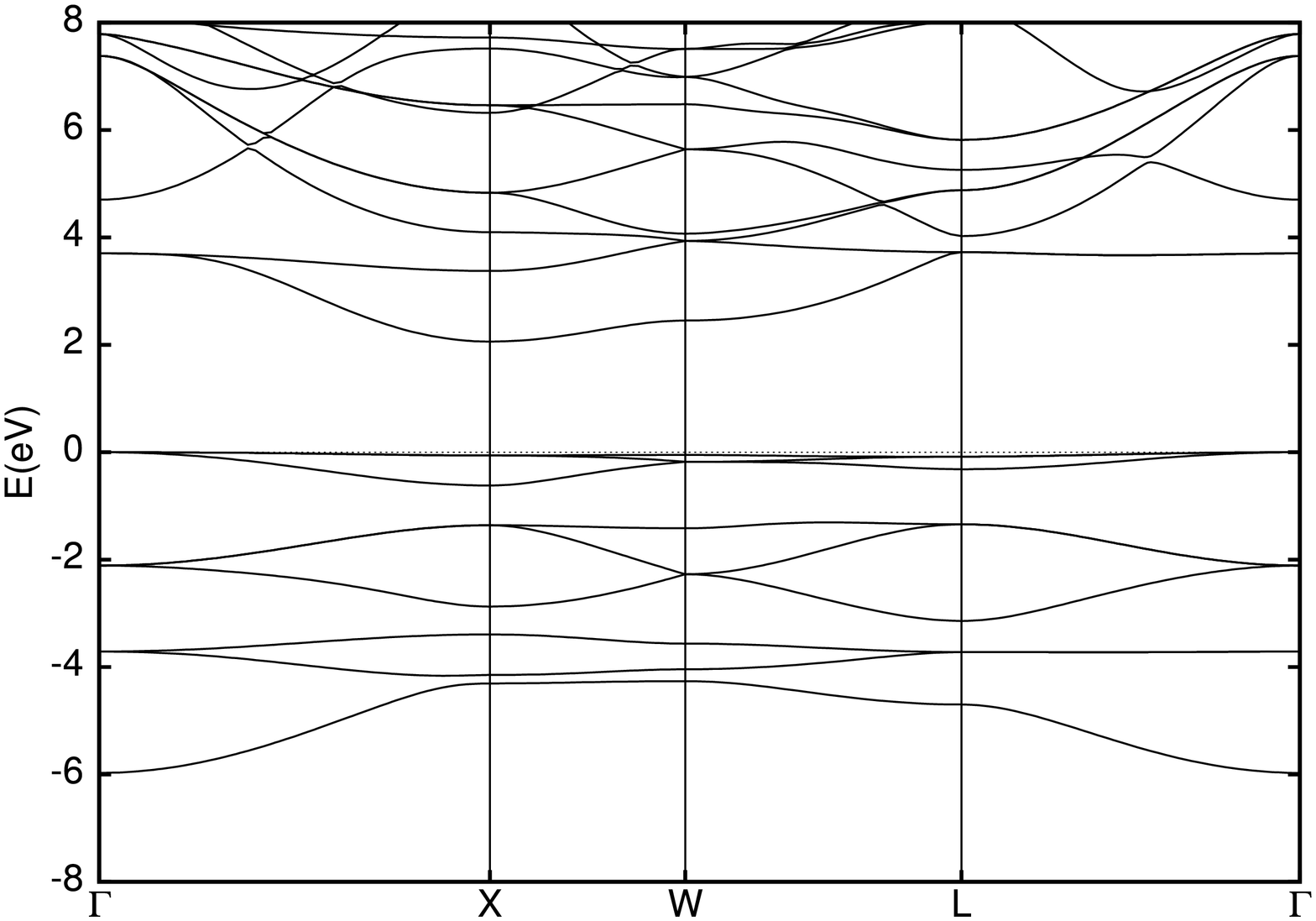,width=0.75\linewidth,angle=270,clip=}}
\vspace{0.1cm}
\caption{Band structure of Mg$_2$RuH$_6$ (top), Ca$_2$RuH$_6$ (middle)
and Sr$_2$RuH$_6$ (bottom), with the relaxed LDA H positions.
}
\label{bands-ru}
\end{figure}

\begin{figure}[tbp]
\centerline{\epsfig{file=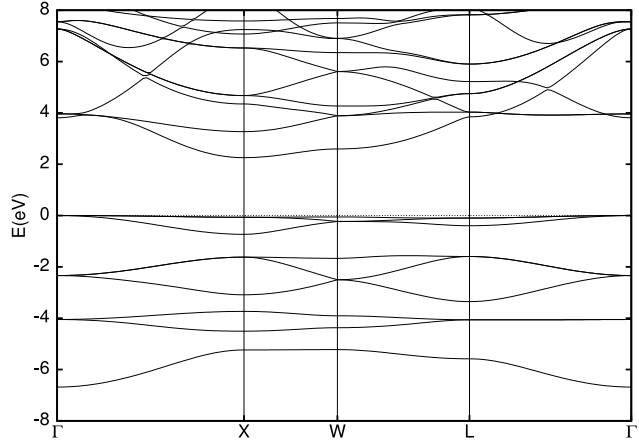,width=0.75\linewidth,angle=270,clip=}}
\vspace{0.1cm}
\caption{Band structure of
Sr$_2$OsH$_6$ with the relaxed LDA H positions.
}
\label{bands-os}
\end{figure}

\begin{figure}[tbp]
\centerline{\epsfig{file=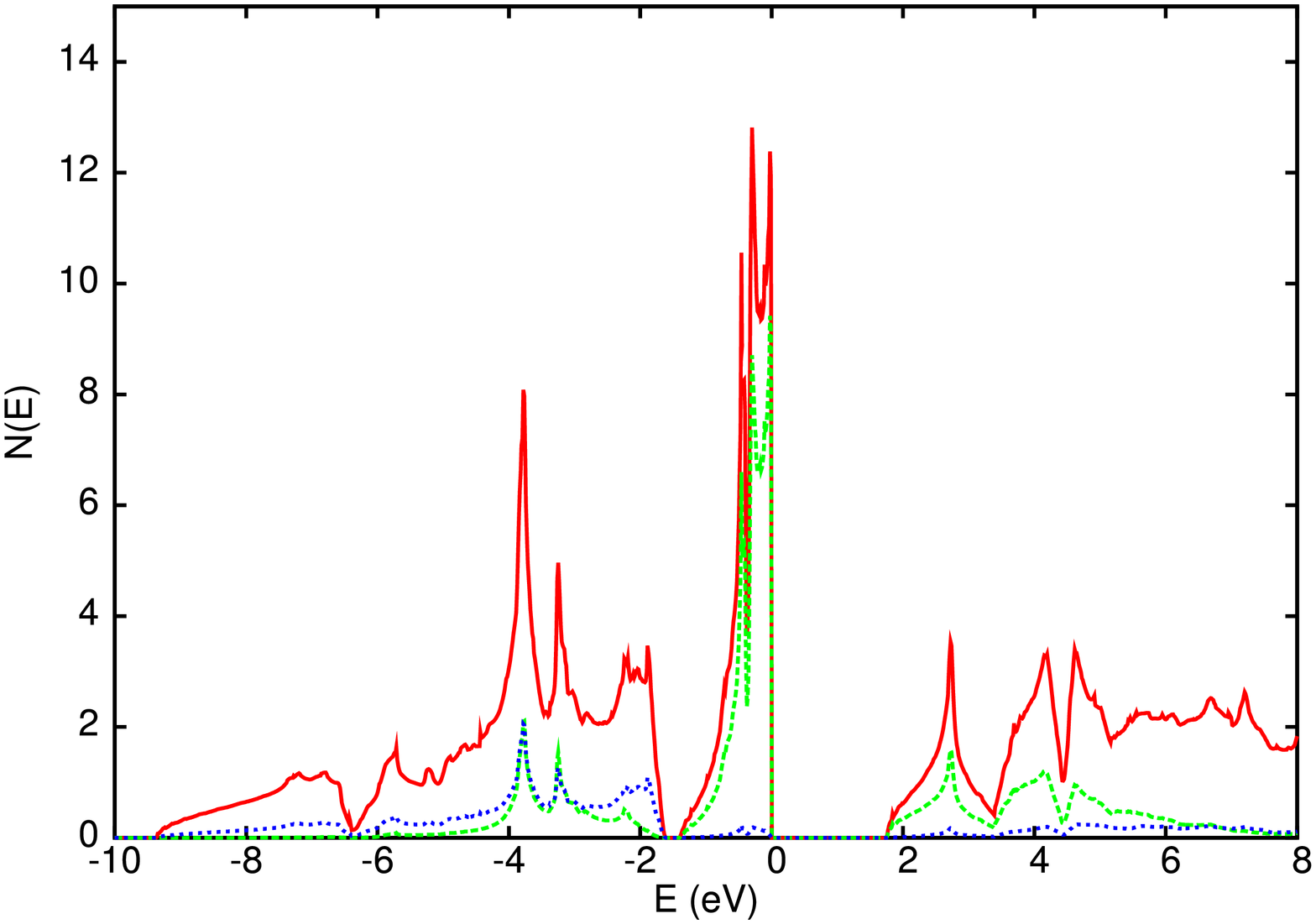,width=0.75\linewidth,angle=270,clip=}}
\vspace{0.1cm}
\centerline{\epsfig{file=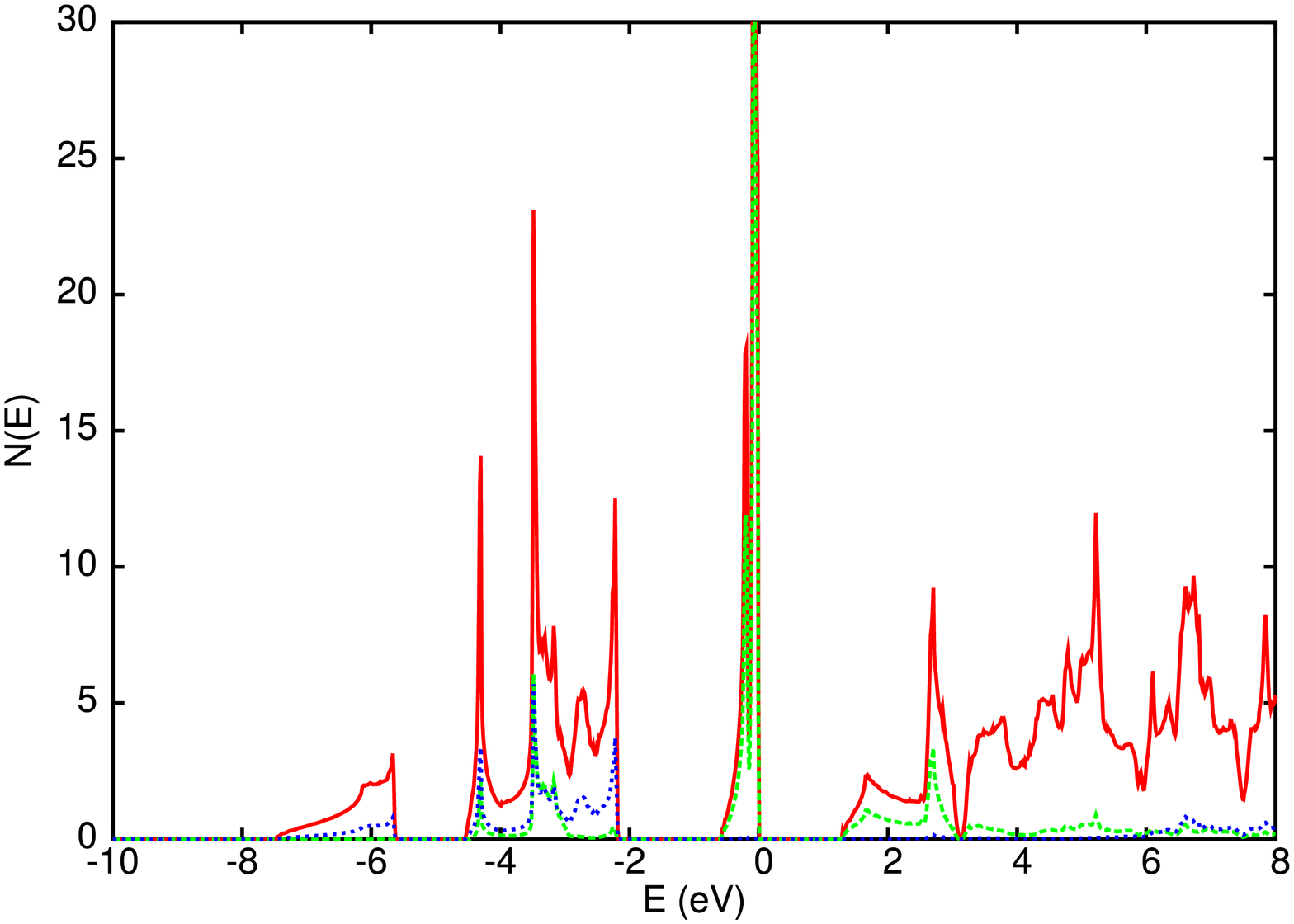,width=0.75\linewidth,angle=270,clip=}}
\vspace{0.1cm}
\centerline{\epsfig{file=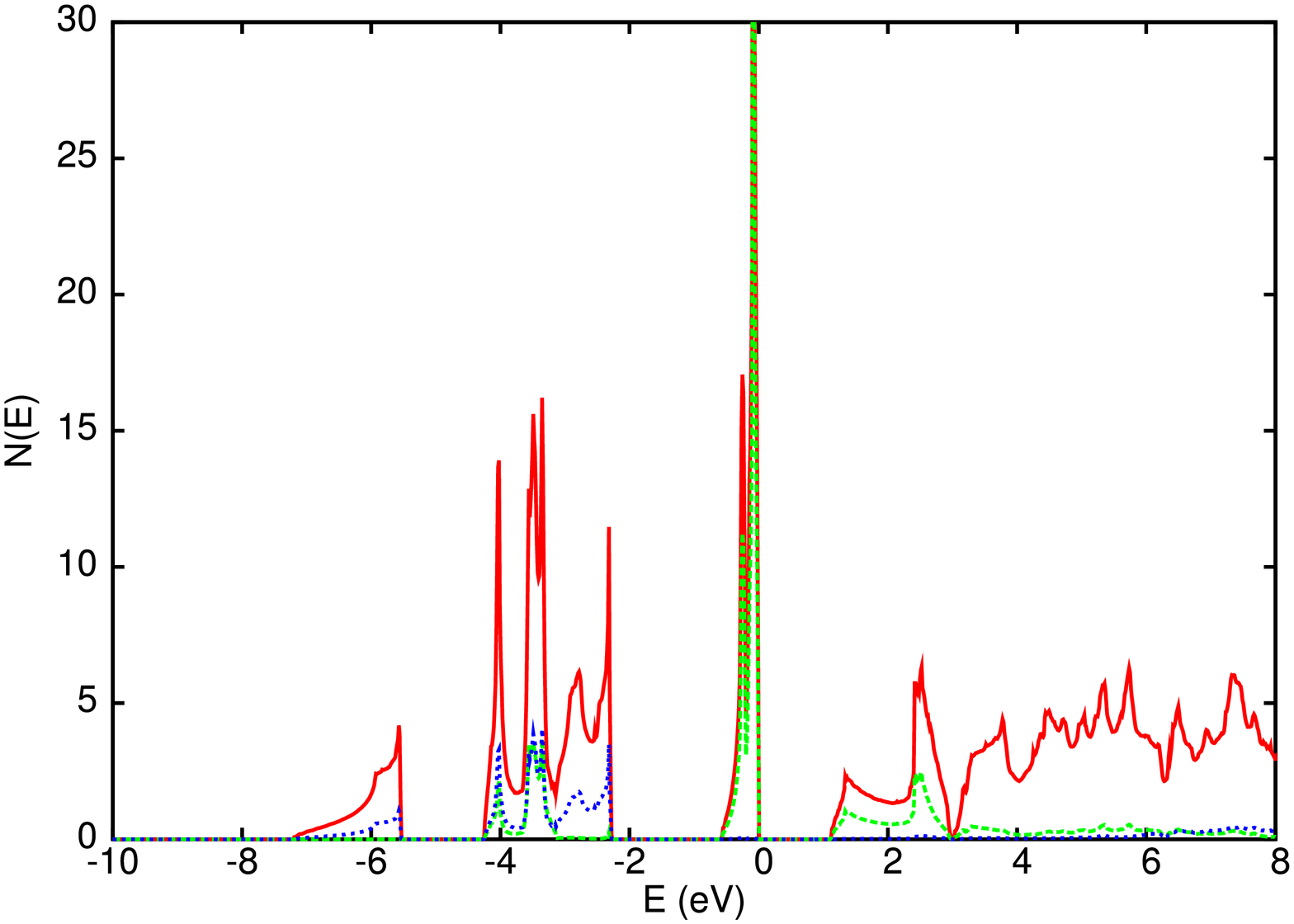,width=0.75\linewidth,angle=270,clip=}}
\vspace{0.1cm}
\caption{(color online) Electronic density of states (heavy red line)
and projections onto the LAPW spheres of Fe $d$ character (dashed green line)
and H $s$ character (dotted blue line) for
Mg$_2$FeH$_6$ (top),Ca$_2$FeH$_6$ (middle)
and Sr$_2$FeH$_6$ (bottom), with the relaxed LDA H positions, on a 
per formula unit both spins basis.
}
\label{dos-fe}
\end{figure}

\begin{figure}[tbp]
\centerline{\epsfig{file=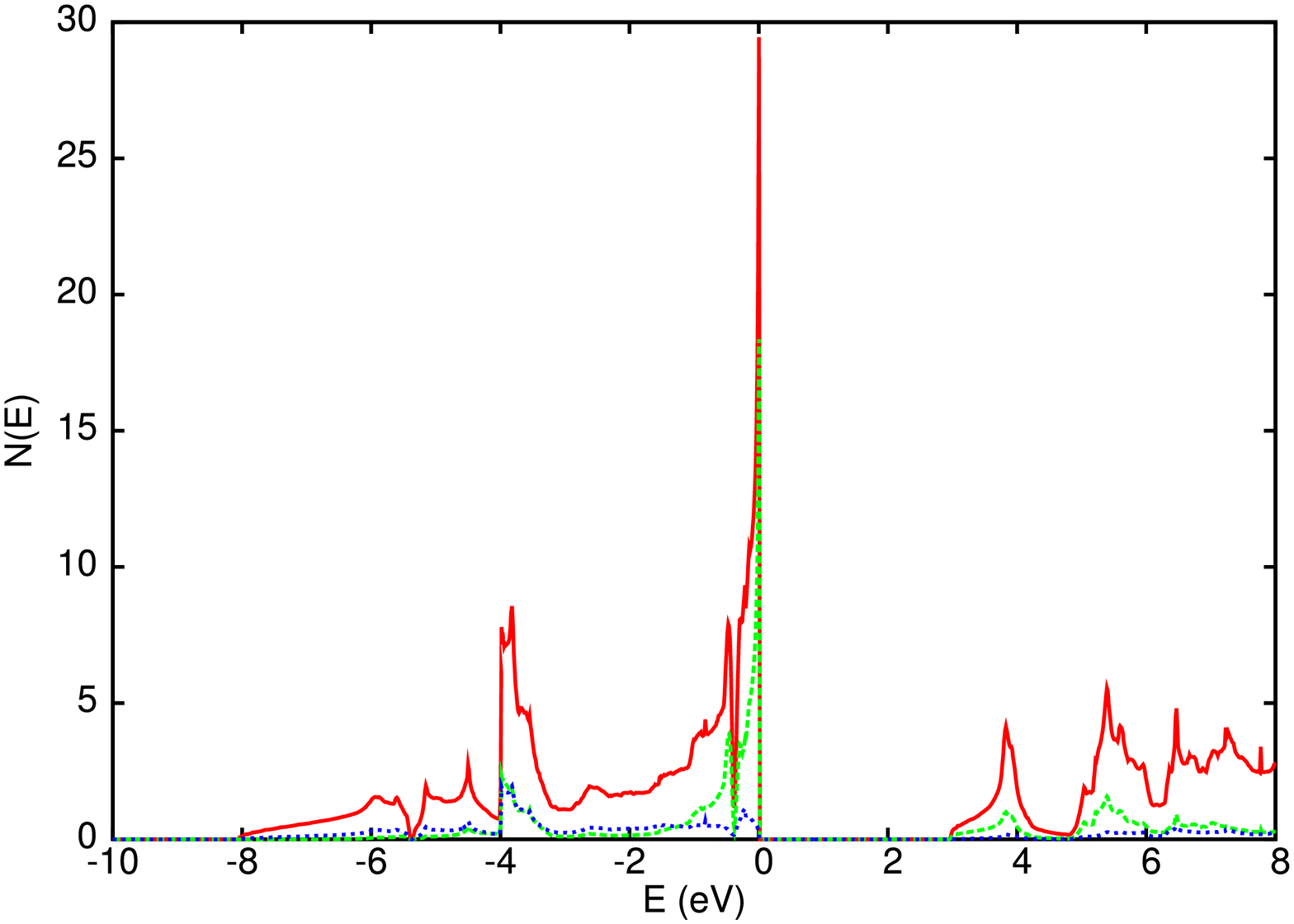,width=0.75\linewidth,angle=270,clip=}}
\vspace{0.1cm}
\centerline{\epsfig{file=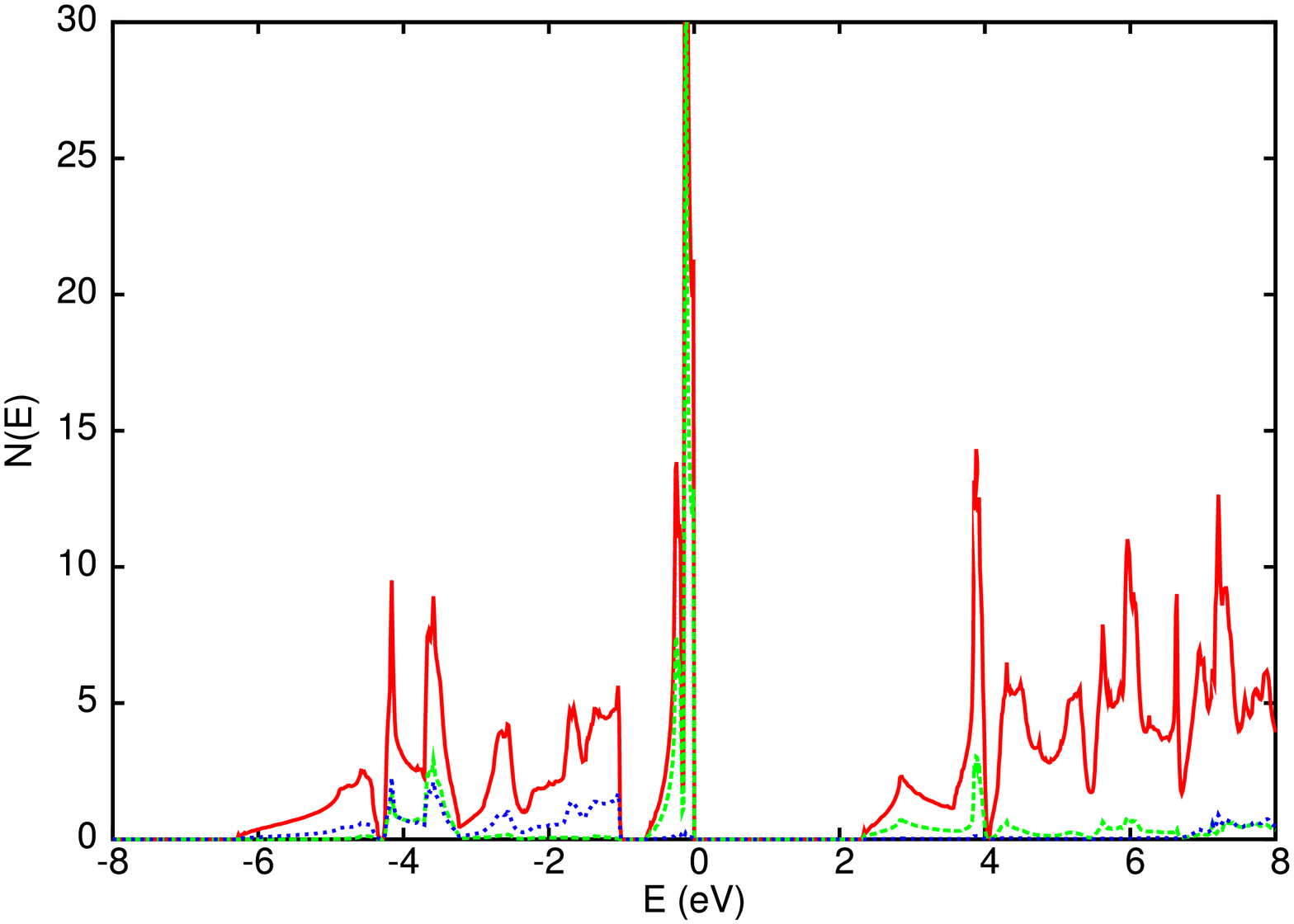,width=0.75\linewidth,angle=270,clip=}}
\vspace{0.1cm}
\centerline{\epsfig{file=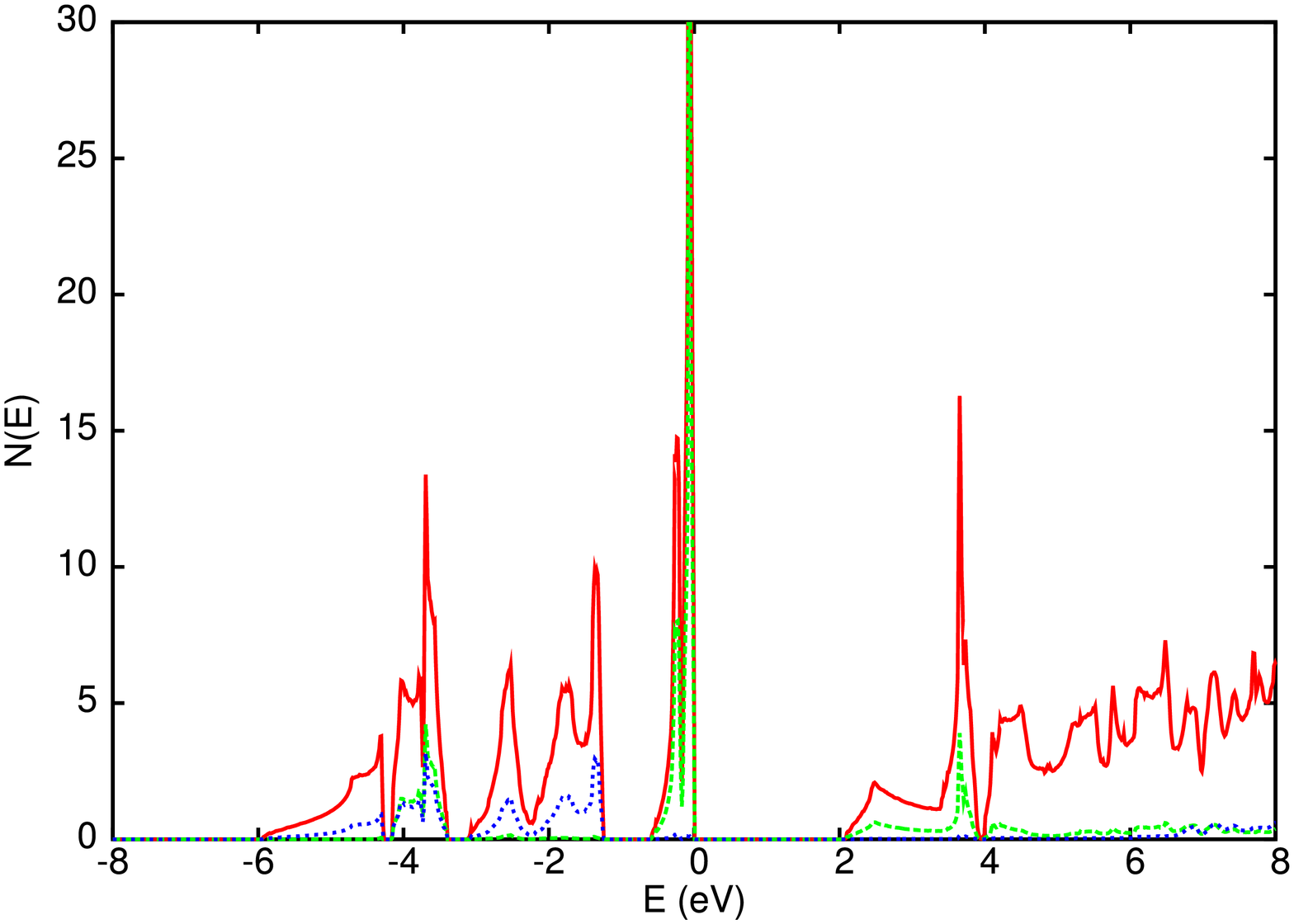,width=0.75\linewidth,angle=270,clip=}}
\vspace{0.1cm}
\caption{(color online) Electronic density of states (heavy red line)
and projections onto the LAPW spheres of Ru $d$ character (dashed green line)
and H $s$ character (dotted blue line) for
Mg$_2$RuH$_6$ (top),
Ca$_2$RuH$_6$ (middle),
and Sr$_2$RuH$_6$ (bottom), with the relaxed LDA H positions, on a 
per formula unit both spins basis.
}
\label{dos-ru}
\end{figure}

\begin{figure}[tbp]
\centerline{\epsfig{file=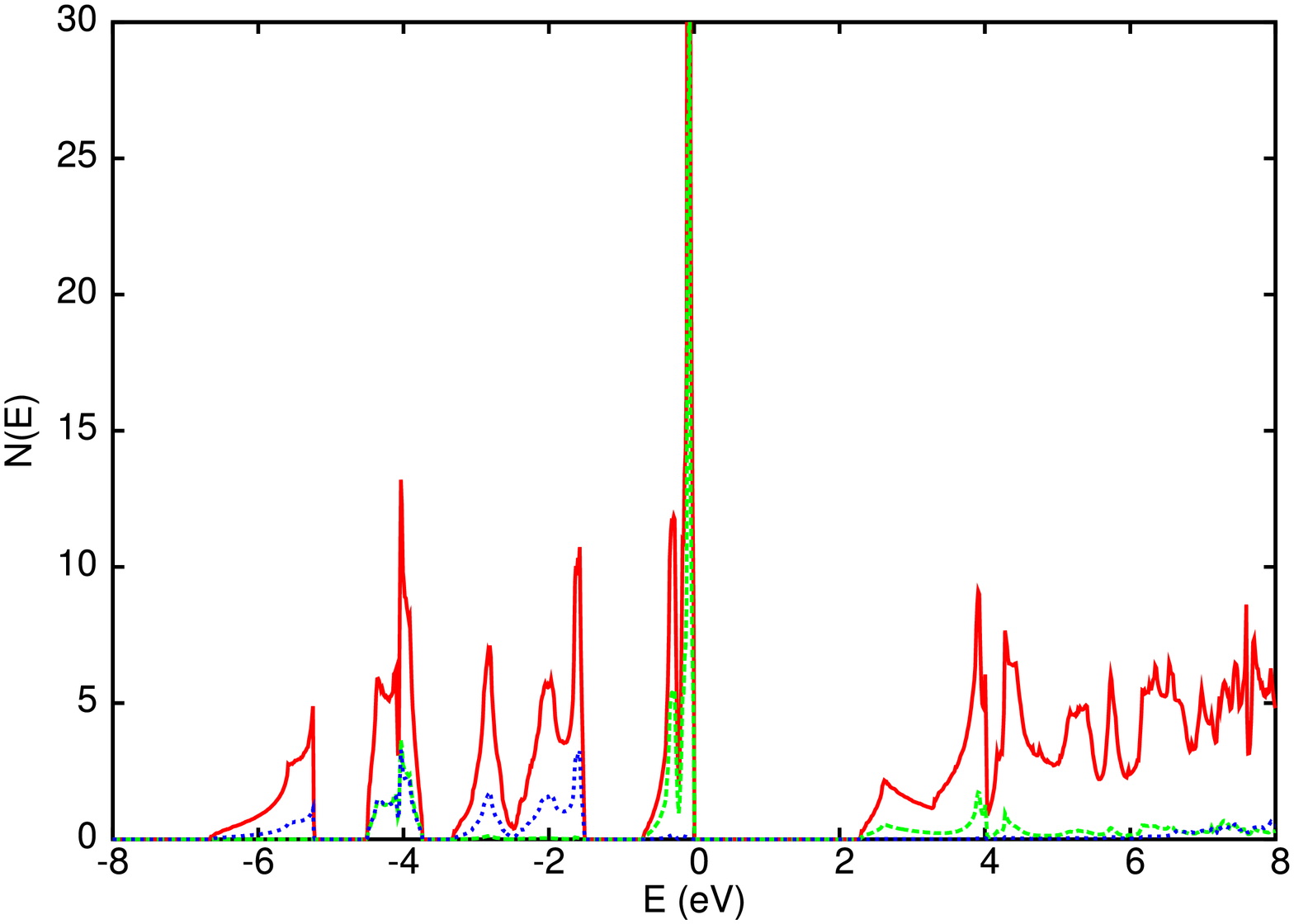,width=0.75\linewidth,angle=270,clip=}}
\vspace{0.1cm}
\caption{(color online) Electronic density of states (heavy red line)
and projections onto the LAPW spheres of Os $d$ character (dashed green line)
and H $s$ character (dotted blue line) for
Sr$_2$OsH$_6$ with the relaxed LDA H positions, on a 
per formula unit both spins basis.
}
\label{dos-os}
\end{figure}

The calculated electronic band structures are shown in
Figs. \ref{bands-fe}, \ref{bands-ru} and \ref{bands-os}.
The corresponding electronic densities of states (DOS) are
in Figs. \ref{dos-fe}, \ref{dos-ru} and \ref{dos-os}.
The electronic structures are qualititively similar to those of
Orgaz and Gupta (Ref. \onlinecite{orgaz}), in that all
the materials are insulating with band
gaps within the transition metal $d$ bands.

The band structure for all the compounds studied consists of 
6 H $1s$ derived bands, holding 12 electrons, followed by
crystal field split transition metal $d$ bands.
The $1s$ band width is largest in the Mg compounds, and
in the case of Mg$_2$RuH$_6$ this is sufficient to yield an
overlap between the $1s$ manifold and the lower $d$ manifold.
In the octahedral
crystal field of the H, the metal $d$ bands separate into
a 3-fold degenerate (6 electrons) $t_{2g}$ manifold and a 
2-fold degenerate (4 electrons) $e_g$ manifold. The 18 valence
electrons then populate the H derived $1s$ bands and the metal
$t_{2g}$ bands; the insulating gap is between the occupied $t_{2g}$
and the unoccupied $e_g$ manifolds.
Since these are gaps within a crystal field split $d$-band,
LDA band gap errors are expected to be small
in the absence of strong correlation effects.
\cite{mattheiss}
However, we note that the gap is between
narrow $d$ bands,
and so a larger gap may be observed both because of optical dipole
selection rules and, in the Fe compounds, correlation effects.

The H $1s$ character of the lowest 6 bands may be seen from
the projections of the DOS onto the LAPW spheres. The small
1.1 $a_0$ H spheres, used here, imply that most of the charge
of H ions will be outside the sphere. A free H$^-$ ion, stabilized
by a Watson sphere (to approximately represent the Madelung field)
of radius 3.01 \AA, would have only 0.45 $e$ (of 2) inside a sphere
of radius 1.1 $a_0$.
Integrating the H $s$ projection of the DOS
over the lowest 6 bands, we obtain from 0.50 $e$ (Ca$_2$RuH$_6$ and
Sr$_2$RuH$_6$) to 0.52 $e$ (Sr$_2$FeH$_6$) inside each H sphere,
not far from this simple ionic view especially
if one allows for a somewhat different breathing.
This is also similar to what was found for NaAlH$_4$. \cite{aguayo}
Thus the basic electronic structure is ionic consisting of H anions and
$D$ and $M$ cations.
These compounds should therefore be viewed as ionic
for understanding the crystal cohesion.

However, from the stand-point of understanding the electronic structure
and H storage properties, covalency is important. The effects of
$M$ -- H hybridization are clearly seen in the electronic structures.
While the bottom 6 bands are essentially H $1s$ bands, they contain the
two
formally bonding H $s$ - $M$ $e_g$ $\sigma$ combinations
(as well as four non-bonding combinations).
The lowest band is the symmetric combination of $s$ orbitals, which
is a non-bonding combination with $M$ $e_g$ states, but is formally
bonding with the nominally unoccuppied $M$ $s$ states. This is
followed by five more H $s$ bands, including the non-bonding and
the formally bonding H $s$ - $M$ $e_g$ $\sigma$ combinations.
In terms of degeneracies, this division into
a ``$s$"-like one fold symmetric band, a
five-fold manifold and the three-fold $t_{2g}$ manifold is
formally like the $s$, $p$, $d$ electron counting of the 18-electron
rule. However, this counting does not correspond to atomic level filling.
Instead, the 2-electron ``$s$" and 10-electron
``$d$" manifolds are actually from combinations of hydrogen
$s$ states, 
and are therefore very different in atomic character from the 6-electron,
``$p$" group, which comes from very weakly
hybridized $M$ $t_{2g}$ bands.

The lowest
conduction bands derive mainly from the corresponding antibonding
combinations and $sd$ derived states associated with the $D$ cation.
The covalency in the [$M$H$_6$]$^{4-}$ units can
be seen in the $M$ $d$ contributions to the H $s$ bands, for example.
The $t_{2g}$ bands show much less hybridization, as expected in
an octahedral ligand environment.
It should be emphasized that the bands show relatively little dispersion,
with the exception of the $D$=Mg compounds,
and that there are generally
clean gaps between the different manifolds (H $s$,
$t_{2g}$ and $e_g$), which implies that weakly
interacting [$M$H$_6$]$^{4-}$ units
may be regarded as the basic building blocks for
understanding the band structure.
The sizable crystal
field splitting of the $M$ $d$ bands underlies
the 18-electron rule in these compounds.
In particular, without
it the $t_{2g}$ and $e_g$ manifolds would
overlap, and then there would be no barrier to adding more than 18 electrons;
the 18 electron rule here is the result of the crystal field splitting of
the metal $d$ levels.
The octahedral geometry, with its large ligand field is energetically
favorable for 18 or fewer electrons. \cite{tet-note}
The substantial crystal field splittings
(as compared to the on-site Hund's coupling, which can be characterized
by a Stoner $I \sim 0.7-0.9$ eV for Fe) \cite{gunnarsson,yamada}
are responsible for the low spin Fe observed in these compounds.

To summarize the results so far, the electronic structure is built
up in the following way in decreasing order of the size of the interactions
involved. (1) Coulomb interactions, particularly the Madelung field,
stabilize an ionic configuration, nominally $D^{2+}M^{4+}$H$_6^{-}$.
This is the main ingredient in the cohesion.
(2) Hybridization between the H $s$ orbitals and the $M$ $e_g$ orbitals
lead to a bonding anti-bonding splitting between these and contribute
to a substantial crystal field splitting between weakly hybridized
occupied $M$ $t_{2g}$ states and unnoccupied $M$ $e_g$ states.
This splitting and the position of the $D$ derived states well above
the $t_{2g}$ energy
underlies the 18 electron rule.
(3) Hopping between the [$M$H$_6$]$^{4-}$ units (presumably mostly
assisted hopping via unnoccipied $D$ $s$ and $d$ states) leads to
band formation. This is reminiscent of some of the oxide double
perovskites, $A_2 MM'$O$_6$, with an inert $M'$, such as
Sr$_2$RuYO$_6$, although in that case the hybridization inside the
[RuO$_6$]$^{7-}$ units is very much stronger than in the present hydrides.
\cite{mazin}

\section{Phonons, Ferroelectricity and Hypothetical
P\lowercase{b}$_2$R\lowercase{u}H$_6$}

The resulting picture of ionic crystal with substantial covalency between
anions and an octahedrally coordinated transition element cation suggests
similarities with double perovskite oxides. Furthermore, the fact that
the latttice contains a large anion (H$^-$) stabilized by the Madelung
field and hybridized with nominally unoccupied transition metal states
further suggests connections with perovskite oxides, particularly
ferroelectrics. In fact, many of the technologically important
ferroelectrics are based on 
perovskites $AB$O$_3$ with mixtures of metal atoms on the $B$ sites 
(these can be disordered or ordered as in {\em e.g.}
double perovskite). Examples include PZT [Pb(Zr,Ti)O$_3$],
PMN-PT [Pb(Mg,Nb,Ti)O$_3$] and PZN-PT [Pb(Zn,Nb,Ti)O$_3$].
In these materials both Pb-O and $B$-O hybridization is important
in the ferroelectricity. \cite{cohen}

In order to further elucidate the relationship to oxides, we calculated
those zone center phonon frequencies of Sr$_2$RuH$_6$
compatible with a rhombohedral $R32$ symmetry, and compare with similar
calculations for hypothetical Pb$_2$RuH$_6$.
This was done at
the experimental lattice parameter of Sr$_2$RuH$_6$,
using the relaxed H position.
This non-centrosymmetric
group would include
the ferroelectric mode, if the material were ferroelectric.
The calculations were 
done by fitting the dynamical matrix to a series of frozen
phonon calculations with small displacements of the various atoms.
This yields 6 three-fold degenerate modes
(plus the three $\omega=0$ acoustic modes).
The frequencies and displacement patterns of the phonon modes are
given in Table \ref{f-table1}.
The 2 \%
difference between the $A_{g}$ frequencies for Sr$_2$RuH$_6$ between
Tables \ref{table1}
and \ref{f-table1} reflect the different approaches
and should be considered indicative of the errors in the fits used
in constructing the dynamical matrix.
The fitting errors can also be
seen in the deviation of the mode character as given in Table \ref{f-table1}
from the $A_g$ character required by symmetry.
For the $A_g$ frequency, the value in Table \ref{table1} should
be considered more reliable because that value was obtained enforcing
the exact mode symmetry, but it should be kept in mind that the LDA
error is likely larger than the difference between the values in Tables
\ref{table1} and \ref{f-table1}.

The highest frequency branches correspond to H motions, as expected.
Of these, the
$A_g$ Raman mode, which corresponds to symmetric breathing of the
RuH$_6$ octahedra, is the stiffest mode, and the next lower mode
also involves modulation of the Ru-H bond lengths. The two intermediate
modes (738 cm$^{-1}$ and 777 cm$^{-1}$) involve 
distortion of the octahedra, which would also yield lower frequency modes
in oxides. The two lowest frequency modes are motions of the Sr
atoms within their cages. The lowest mode is the antisymmetric motion of
the Sr, which is not ferroelectric.
The frequencies of these Sr motions are
compatible with the frequencies
of the shearing modes that modulate Sr-H distances
when the mass difference is accounted for.

This pattern of phonon modes
is quite different from what would occur in an oxide near ferroelectricity.
In that case, there would be a low frequency cooperative mode. This would
consist
of a distortion where the cations move relative
to the O atoms comprising the octahedra, reflecting the high polarizability
of the octahedra softened by covalent interactions. \cite{cohen,singh-fe}
Here, the mode corresponding to motion of the Ru with respect to the
H is at high frequency (1433 cm$^{-1}$) and the lower frequency Sr
derived modes have only a small component of Ru motion relative to the
H octahedra.

We repeated the calculations for the hypothetical compound
Pb$_2$RuH$_6$. In perovskite and double perovskite oxides, Pb typically
can be substituted for Sr. The Pb compounds typically have unit cell
volumes very close to the Sr analagues, but are more likely to be
ferroelectric because of Pb-O covalency (e.g. PbTiO$_3$ vs. SrTiO$_3$).
Since the experimental lattice parameter of Pb$_2$RuH$_6$ is unavailable,
we used the value for Sr$_2$RuH$_6$. This choice also makes comparison of
the two systems more direct.
LDA relaxation of the lattice parameter
yielded a value 1.8\%
smaller than this, but considering the usual underestimate of lattice
parameters in the LDA,
we do not consider the LDA value to be more reliable than
the use of the Sr$_2$RuH$_6$ value. \cite{pb-lat}
The Pb$_2$RuH$_6$ modes
(Table \ref{f-table2})
are
qualitatively like those of Sr$_2$RuH$_6$, except that the low frequency
Pb modes are shifted down in frequency.
This shift is, however, larger
than can be accounted for by the mass difference, and the modes are
reversed in order. The lowest mode is now the symmetric mode
and it is at zero frequency to within the precision of the present
calculations.
We also calculated the energy as a function of the rotation of the
RuH$_6$ octahedron. Such rotational degrees of freedom compete with
ferroelectricity in ferroelectrics such as Pb(Zr,Ti)O$_3$. \cite{fornari}
Here, these modes are stable. We obtain frequencies of 576 cm$^{-1}$ and
490 cm$^{-1}$, for Sr$_2$RuH$_6$ and Pb$_2$RuH$_6$, respectively.
This suggests Pb$_2$RuH$_6$ as a candidate
ferroelectric
hydride. We found a similar result for hypothetical
Be$_2$FeH$_6$. In this case, we
obtained a slightly unstable mode of Be character, with a ferroelectric
displacement pattern, and stable rotational and antiferroelectric
modes.
If these compounds are made, the possibility of ferroelectricity
should be investigated, {\em e.g.} by low
temperature structural studies and temperature dependent dielectric
measurements.

\section{Energetics and Zero Point Effects}

In order to better understand the stability of these compounds, we performed
calculations of the formation enthalpies by comparison
of the total energies with those of decomposition products.
Specifically, we did calculations for
the elements, Mg,Sr,Ca,Fe,Ru and Os
in their bulk metallic form
(in the LDA at the experimental lattice parameters,
parallel to the calculations done for the hydrides,
including ferromagnetism for Fe),
the H$_2$ molecule (relaxed, in the LDA) and MgH$_2$, CaH$_2$ and
SrH$_2$ (using experimental lattice constants, but relaxed atomic
positions). In addition calculations were done for elemental Be and
the intermetallic phase Be$_2$Fe, which is the expected decomposition
product of the hypothetical phase Be$_2$FeH$_6$ (see below).

For the H$_2$ molecule we used a cubic supercell of lattice parameter
4.5 \AA. This yielded an LDA energy of -2.288 Ry, and bond length
of 0.765 \AA, and bond stretching vibrational frequency of 4217 cm$^{-1}$.
These results are in good agreement with previous LDA calculations.
For example, Patton and co-workers report a vibrational frequency
of 4188 cm$^{-1}$ and bond length of 0.765 \AA. \cite{patton}
The H$_2$ zero point energy obtained from the LDA frequency is 25.2
kJ/mol. This is a substantial number, which underscores the fact
\cite{wicke,tao,miwa}
that zero point effects need to be considered in the thermodynamics of
hydride formation. Here we neglect metal modes, and consider only
the H contribution to the zero point energies, which we write as
$3\hbar\bar{\omega}$ per H$_2$ unit, where $\bar{\omega}$ is an average
H vibrational frequency. The effective $\bar{\omega}$ for H$_2$
is 703 cm$^{-1}$, so for hydrides with $\bar{\omega} > 703$ cm$^{-1}$,
zero point motion will reduce the formation energy, and the
corresponding deuterides and tritides will form more easily than the hydrides,
while for materials with $\bar{\omega} < 703$ cm$^{-1}$, the converse will
be true. \cite{wicke}
At least in principle, this difference can be used to obtain
the average H frequency from experimental formation energies of hydrides
and deuterides, but to our knowledge this has not been done for these
materials.

In order to estimate $\bar{\omega}$ for the compounds considered
here, we performed LDA calculations for selected distortions and assumed
that the H behaves in an Einstein-like way. In particular, for MgH$_2$,
we displaced a single H in the unit cell (which contains 4 equivalent H atoms)
along the three principal directions in its cage and averaged the
resulting frequencies to obtain an average frequency (the principal
directions relative to the lattice are [1,$\bar{1}$,0], [1,1,0]
and [0,0,1]). For CaH$_2$ and SrH$_2$, we used the average of the four
full symmetry Raman modes that are H derived (there are two other such
modes associated with metal motion). For the K$_2$PtCl$_6$ structure
hydrides, we used the average of the highest two
$\Gamma$-point modes consistent with rhombohedral symmetry (the highest of
these is the full symmetry Raman breathing mode)
to obtain an effective $B$-site metal -- H bond stretching force constant
(which contributes 1/3 of the modes) and obtained the effective force
constant for the
other two thirds of the modes by averaging the two other H derived modes
from the rhombohedral symmetry and the octahedral rotation mode.
As a test, we also calculated an average Einstein frequency for
Mg$_2$FeH$_6$ by displacing a single H (of the six in the unit cell)
perpendicular to the Fe-H ``bond", and along it, similar to the
procedure that was followed for MgH$_2$. This yielded a shear frequency
of 727 cm$^{-1}$ and a stretch frequency of 1828 cm$^{-1}$, for an
average $\bar{\omega}$=1094 cm$^{-1}$, in fortuitously good agreement with the
estimate of 1089 cm$^{-1}$, made as above (Table \ref{table3}).
In all cases, the averages are arithmatical averages of frequencies as
is appropriate for the zero point energy.

$\alpha$-MgH$_2$ has a tetragonal structure
(spacegroup $P4_2/mmm$) with one H coordinate, $x$. We obtain
$x$=0.3046, in agreeement with the recent neutron measurement of
Bortz and co-workers, \cite{bortz} who obtained $x$=0.3040, and
an LDA calculation by Yu and Lam, \cite{yu} who also obtained $x$=0.304.
Moving a single hydrogen in the unit cell of MgH$_2$, we obtained
frequencies of 1277 cm$^{-1}$ along [1,1,0] (bond stretching)
592 cm$^{-1}$ along [1,$\bar{1}$,0] (bond bending) and 993 cm$^{-1}$
along [0,0,1] (mixed). The bond stretching ``Einstein" frequency is near
that of the full symmetry Raman mode, which we obtain at 1301 cm$^{-1}$
and which is also of bond stretching character. This supports the simple
Einstein-like approach used.

Good agreement
with neutron diffraction results \cite{andresen,brese}
is also obtained for the internal structural parameters of
CaH$_2$ and SrH$_2$, which occur in an orthorhombic $Pnma$
structure, as given in Table \ref{table2}.
As mentioned, the full symmetry Raman frequencies obtained from
this relaxation, were used to construct the H frequency for the
zero point contribution to the enthalpy of these compounds.

LDA formation energies are given in Table \ref{table3}.
The formation energies of MgH$_2$, CaH$_2$ and SrH$_2$
are in excellent agreement with experiment and also in
good agreement with previous LDA calculations.
\cite{smithson}
The formation energy of Mg$_2$FeH$_6$ is the best studied of the
K$_2$PtCl$_6$ hydrides, and is reported as -98 kJ/mol (Ref.
\onlinecite{didisheim}), -86 kJ/mol (Ref. \onlinecite{konst}),
and -77.4 kJ/mol (Ref. \onlinecite{bogdanovic}). The calculated
LDA energy of -133 kJ/mol is significantly larger and this
difference would seem to be at the high end of the normal range of LDA
errors,
especially considering the good agreement with experiment
for MgH$_2$, and FeAl, where we obtain
agreement with experiment to within 5 kJ/mol of Fe. \cite{e-note}
Besides LDA errors, the most likely source of error
is the crude method that we used to obtain an average phonon frequency.
However, even if the average phonon frequency were 250 cm$^{-1}$ higher
than our estimate, which we think is unlikely, the calculated enthalpy
would shift by only 9 kJ/mol of H$_2$.

One possibility is that some of the difference is experimental in origin,
related to the possible existence of some stable
hydride among the decomposition products,
which would then stabilize the products relative to Mg$_2$FeH$_6$
and therefore would lower the formation energy as measured by the
decomposition.
In any case, we do find certain trends. First of all,
Mg$_2$FeH$_6$ is by far the least stable of the K$_2$PtCl$_6$ hydrides
studied.
However, this is connected with the fact that MgH$_2$
is much less
stable than SrH$_2$ or CaH$_2$. If one considers formation via
$ 2 D {\rm H}_2 + {\rm H}_2 + M \rightarrow D_2M{\rm H}_6 $,
then this heat of formation is largest for Mg$_2$FeH$_6$
as might be expected from an ionic picture.
Secondly, the formation energy per H$_2$ is
significantly larger for Mg$_2$FeH$_6$
than for MgH$_2$. This implies that under thermodynamic conditions,
without some very unusual entropy contribution,
the decomposition should proceed directly to H$_2$ and the elements,
or, at a bare minimum,
if there is an intermediate hydride phase, it should not be MgH$_2$.
The formation energies of the Ca and Sr compounds on the other hand
are very close to those of CaH$_2$ and SrH$_2$, so depending on
the conditions, those decompositions may very well proceed via
an intermediate $M$ + $D$H$_2$.

Similar calculations were done for the other compounds in order
to estimate the H zero point energy, but these were at a lower level
of convergence in the fitting of the dynamical matrix.

\section{Stability, Bonding and Implications}

We now speculate about possible implications of our results
for hydrogen storage.
First of all, we note that the cohesion is ionic, and that it is
the Madelung field that stabilizes the [$M$H$_6$]$^{4-}$ units.
Changes in the Coulomb potential then ought to strongly affect the
bond lengths in these units as well as their stability. This is
already apparent in the values of $x_{\rm H}$ of Table \ref{table1},
which show substantial changes in the $M$-H bond lengths as the
lattice parameter is changed by $D$ site substitution (note that the
octahedra are not connected so that they need not breath with the lattice).
This implies significant tunability in the properties with substitutions.
Secondly,
the ionic stabilization of the lattice implies that mixed $M$ cation
substitutions should be possible and that the octahedral coordination
of the metal atoms will be preserved in them. For example,
if the partial or full substution
Os $\rightarrow$ ${{1}\over{2}}$ Re + ${{1}\over{2}}$ Ir
could be made, its structure is expected to feature ReH$_6$ and IrH$_6$
octahedra, rather than different Re-H and Ir-H coordinations on this lattice.
However, it is unclear if any of these substitutions can be made, and even
if they can it is unclear whether they will be beneficial.
Finally, we note that
the fully hydrided compound can therefore be
destabilized by driving the transition metal
$d$ states up in energy via the Madelung potential, 
if all other things were equal. One way would be to substitute some
fraction of the Mg with a monovalent cation if one can be made to
enter the lattice.

However, the stability is relative to the decomposed products, and
it is clear from the calculated energetics that these play a major
role. For example, Mg$_2$RuH$_6$ and Mg$_2$FeH$_6$ are the least
stable compounds relative to decomposition into elemental Mg and Ru or Fe,
but they are the most stable with respect to a hypothetical intermediate
MgH$_2$ + Fe/Ru.
As mentioned, 
this suggests that under normal conditions
Mg$_2$FeH$_6$ and Mg$_2$RuH$_6$ decompose into
H$_2$ and elemental metals without any MgH$_2$ intermediate, consistent
with the observation of Bogdanovic and co-workers. \cite{bogdanovic}

\section{Reducing the Formation Energy: Hypothetical
B\lowercase{e}$_2$F\lowercase{e}H$_6$}

Considering the trend in the energetics with respect to the alkaline
earth element, one possibility for obtaining a lower formation energy
would seem to be replacement of Mg by Be. This would seem especially
likely considering the properties of Be metal, which include strong bonding
that would compete with the formation of hydride phases.
In order to check this trend
we performed calculations for hypothetical Be$_2$FeH$_6$ to obtain its
formation energy. Since the compound is hypothetical, we obtained
the lattice parameter by relaxation in the LDA. The calculated structure
has a lattice parameter of 5.65 \AA, a H internal coordinate of $x$=0.2648,
and a corresponding full symmetry Raman phonon frequency of 2233 cm$^{-1}$.
No doubt the LDA underestimates the lattice parameter slightly, as is
typical. In any case, with the LDA structure, we obtain a static
formation enthalpy of -37 kJ/mol H$_2$
with respect to elemental products.
This confirms the conjecture
that Be would lead to lower binding energies. However,
while this energy suggests that Be$_2$FeH$_6$ would be an interesting
hydride phase, it neglects the fact that unlike Mg, Be forms compounds
with Fe. In particular, FeBe$_2$ is a known intermetallic compound and
would compete with the hydride phase. We calculated the enthalpy of
formation of FeBe$_2$ (details are in the next section) and find
-87 kJ/mol.
The relevant energy for the stability of the
hydride Be$_2$FeH$_6$ is therefore not the formation enthalpy from
the elements, but from the intermetallic, BeFe$_2$, which, with H$_2$,
would be the product of the decomposition.
Relative to decomposition into the elements, the existence of the intermetallic
then results in a shift of the formation enthalpy
of Be$_2$FeH$_6$ by 29 kJ/mol on a per H$_2$ basis, to yield
-8 kJ/mol. The average phonon frequency, determined as for the other
K$_2$PtCl$_6$ hydrides, discussed above, is 1172 cm$^{-1}$, yielding a
zero point correction of +17kJ/mol H$_2$, placing the calculated enthalpy
including zero point at +9 kJ/mol.
Thus, it is likely
that Be$_2$FeH$_6$ is marginally unstable with respect to
decomposition into FeBe$_2$, and therefore will
only be formed under high pressure or by chemical routes.

\section{Electronic Structure and Magnetism in
F\lowercase{e}B\lowercase{e}$_2$}

FeBe$_2$ is an interesting hard magnetic material. In particular,
it has a relatively low density, high anisotropy and a very
high Curie temperature, $T_C$ of 880 K. \cite{str-febe2,jesser,samata}
Experimentally, FeBe$_2$ crystallizes in the hexagonal MgZn$_2$ structure
(spacegroup $P6_3/mmc$, No. 194)
and has magnetization, $M \approx 1.95 \mu_B$/Fe.
Since we are not aware of previous first principles studies of this
material, we briefly summarize our results for the electronic structure of
this compound.

The reported lattice parameters are $a$=4.215 \AA, and 
$c$=6.853 \AA. \cite{str-febe2}
The unit cell contains four formula units. The Be atoms
are on sites $2a$ (0,0,0) and $6h$ ($x$,2$x$,1/4), while
the Fe atoms are on $4f$ (1/3,2/3,z). Experimental
values of the two internal parameters are not available from experiment,
so they were found by structural relaxation in the LDA.
We find $x$= 0.8294 and $z$=0.061. The calculated LDA spin magnetization
is 1.76 $\mu_B$/Fe and the magnetic energy is 0.237 eV/Fe.
This is only $\sim 3kT_C$ suggesting some itinerant character.

\begin{figure}[tbp]
\centerline{\epsfig{file=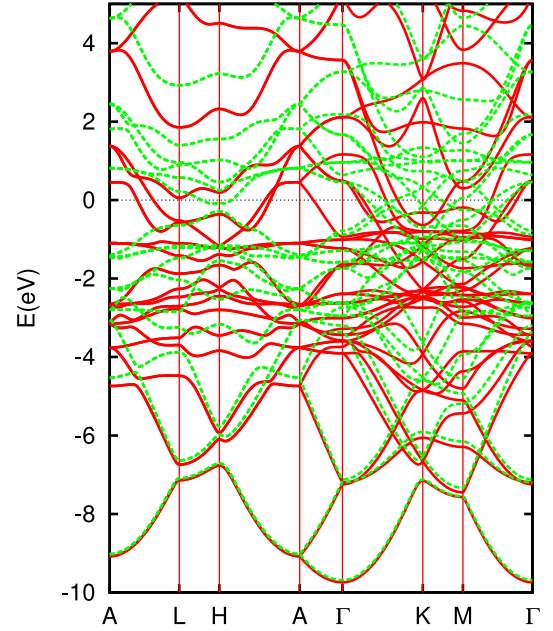,width=1.10\linewidth,angle=270,clip=}}
\vspace{0.2cm}
\caption{(color online) Band structure of ferromagnetic FeBe$_2$.
Majority (minority) spin bands are given by solid red (dashed green)
lines. The Fermi energy is at 0 eV.
}
\label{befe-bands}
\end{figure}

\begin{figure}[tbp]
\centerline{\epsfig{file=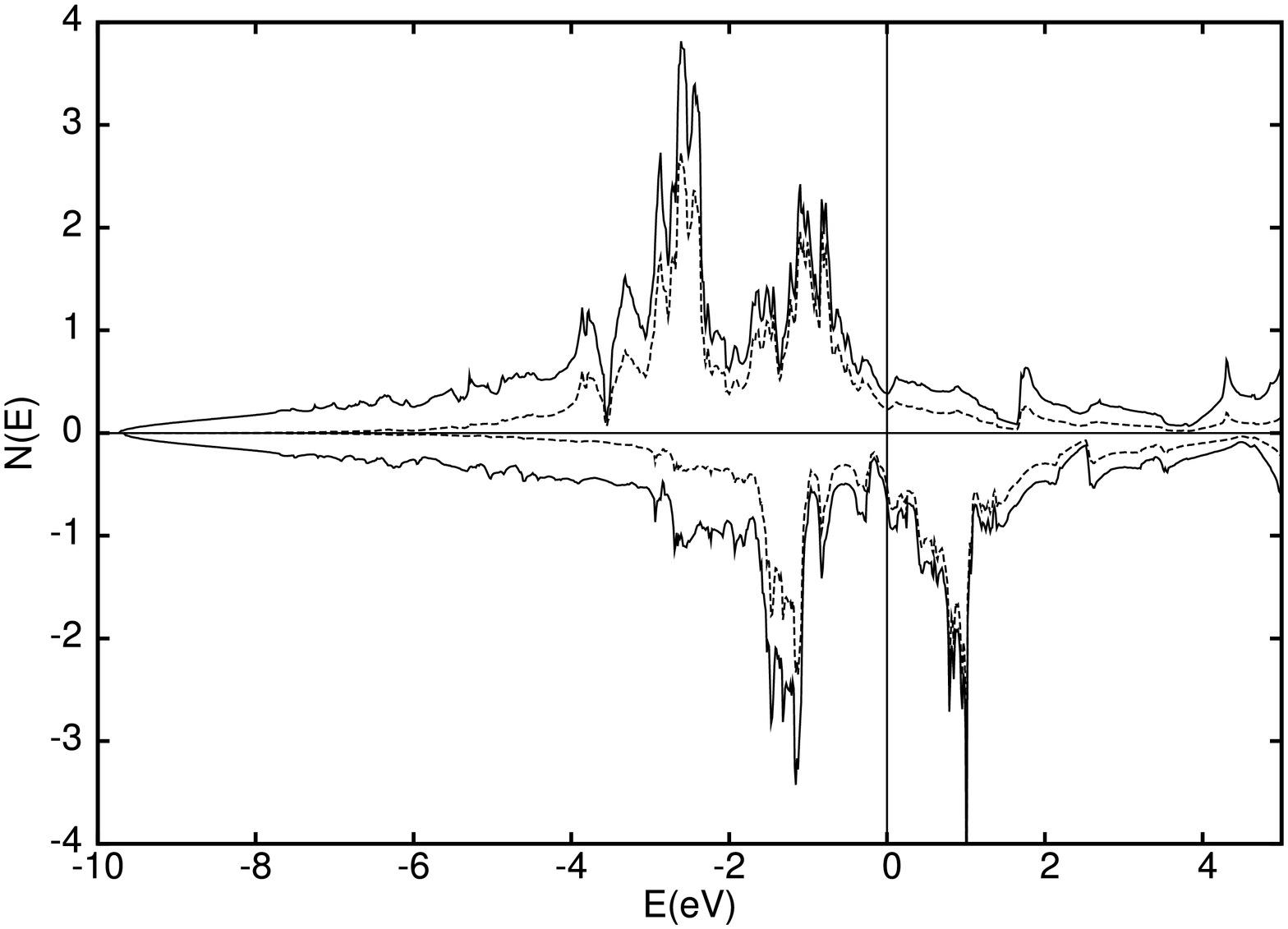,width=0.75\linewidth,angle=270,clip=}}
\vspace{0.1cm}
\caption{Electronic density of states (solid) and Fe $d$ projection onto
the LAPW sphere (broken, radius 2.1 $a_0$) for FeBe$_2$ on a per
per formula unit basis. Spin-up is above the 0 and spin down is shown as
negative. The Fermi level is at 0 eV.
}
\label{befe-dos}
\end{figure}

The local spin density approximation (LSDA) band structure and density
of states are given in Figures \ref{befe-bands} and \ref{befe-dos}.
The band structures shows narrow crystal field split Fe 3$d$ bands
on top of a broad manifold of free electron like Be $sp$ derived bands.
These Be derived bands are weakly polarized, opposite to the Fe polarization,
similar to the case of YFe$_2$, for example. \cite{mohn,singh-y}
The majority spin Fe 3$d$ bands are fully occupied, while the Fermi
energy falls in the crystal field gap between the $t_{2g}$ and $e_g$
manifolds in the minority spin.
This yields two minority spin $e_g$ holes
per Fe, and explains the $\sim$ 2 $\mu_B$/Fe magnetization. Relative
to $bcc$ Fe, there is a transfer of Fe $s$ character to the Be derived
bands, and a back transfer of charge to give effectively neutral Fe,
with eight 3$d$ electrons. This is consistent with the picture discussed
by Jesser and Vincze based on experimental susceptibility and Mossbauer
measurements. \cite{jesser}
This pseudogapped band structure
yields a relatively low density of states at the Fermi energy
in both spin channels, $N_\uparrow(E_F)$=0.39 eV$^{-1}$
and $N_\downarrow(E_F)$=0.67 eV$^{-1}$ on a per formula unit basis.

The calculated formation energy of FeBe$_2$ is -87 kJ/mol on a per formula
unit basis.
As a test of our approach, we also calculated the formation energy of
FeAl in the same way. The result was -77 kJ/mol(FeAl),
which is in good agreement with the experimental value
of -72.6 kJ/mol. \cite{e-note,breuer}
This suggests that the error in the formation energy of FeBe$_2$
is likely in the range of 5 kJ/mol.

\section{Summary and Discussion}

The present density functional calculations show that the
K$_2$PtCl$_6$ hydrides are ionic compounds, with some covalency.
The 18 electron rule is a consequence of ligand field effects on the
transition metal site.
This type of ionic character suggests the possibility of ferroelectricity
in related hydrides. We find that the hypothetical compounds, Pb$_2$RuH$_6$
and Be$_2$FeH$_6$ are on the borderline of ferroelectricity,
and should be investigated in this context, if they can be made.
Further, the ionic character stabilizes H anions, and Fe cations,
which is why Fe participates in hydride formation, although Fe metal
is not a hydride former.
The ionic character implies a certain degree of tunability of the
properties of these hydrides, which may allow adjustment of the
thermodynamics. However, since Fe and Mg do not form intermetallic
compounds, it is likely that the properties of
Mg$_2$FeH$_6$ cannot be made better for vehicular
H storage than those of MgH$_2$, since MgH$_2$ will be a competing phase.

One way to reduce the stability of the hydride
without facing this limitation
would be to focus on the stability of decomposition products.
One possibility would be to explore
minor additions, $X$ that are soluble in and stabilize
an Fe-Mg-$X$ intermetallic. These need
not enter the hydride lattice, provided that they are available e.g.
on hydride particle surfaces to promote the decomposition and provide
sufficient enthalpy via the formation of the intermetallic.
This may be the most promising avenue for modifying Mg$_2$FeH$_6$
for hydrogen storage applications.
As far as we are aware, the solubility of Mg in FeBe$_2$ is not known.
However, if Fe-Be-Mg intermetallics are stable, the present results suggest
that the addition of Be to Mg$_2$FeH$_6$ may lead to lower stability,
which if not for the toxicity of Be, would be favorable for applications.

\acknowledgements

We are grateful for helpful discussions with
P. Dantzer, M.R. Pederson and K. Yvon.
DJS thanks the University of Paris-Sud for their hospitality,
which made this work possible.
We also thank the Institut du Developpement et des Ressources en 
Informatique Scientifique (IDRIS) for a grant of computer time.
Work at the Naval Research Laboratory is supported by ONR.

\begin{table}[tbp]
\caption{LDA and experimental H positions ($x_{\rm H}$(LDA) and
$x_{\rm H}$(EXP), respectively),
fully symmetric $A_{g}$ Raman frequency, $\omega$,
and band gap, $E_g$.}
\begin{tabular}{lcccc}
  & $x_{\rm H}$(LDA) & $x_{\rm H}$(EXP) & $\omega$(cm$^{-1}$) & $E_g$(eV)  \\
\hline
Mg$_2$FeH$_6$  & 0.2412 & 0.2420 [\onlinecite{didisheim}~] &
                                                    1923 & 1.73 $X$-$X$ \\
Ca$_2$FeH$_6$  & 0.2257 & 0.2300 [\onlinecite{huang}~] &
                                                    1759 & 1.27 $X$-$X$ \\
Sr$_2$FeH$_6$  & 0.2178 & & 1694 & 1.09 $\Gamma$-$X$ \\
Mg$_2$RuH$_6$  & 0.2527 & 0.2524 [\onlinecite{huang}~] &
                                                    1977 & 2.93 $X$-$X$ \\
Ca$_2$RuH$_6$  & 0.2359 & & 1837 & 2.29 $X$-$X$ \\
Sr$_2$RuH$_6$  & 0.2254 & 0.223 [\onlinecite{moyer}~]
                                             & 1778 & 2.06 $\Gamma$-$X$ \\
Sr$_2$OsH$_6$  & 0.2262 & 
                                             & 1893 & 2.26 $\Gamma$-$X$ 
\end{tabular}
\label{table1}
\end{table}

\begin{table}[tbp]
\caption{LDA and experimental atomic positions ($x$(LDA),
$z$(LDA) and
$x$(EXP), $z$(EXP) for $Pnma$ CaH$_2$ and SrH$_2$. $y$=1/4
for all atoms in this structure.
}
\begin{tabular}{lddddc}
 & $x$(LDA) & $z$(LDA) & $x$(EXP) & $z$(EXP) & \\
\hline
CaH$_2$ Ca & 0.2380 & 0.1100 & 0.2378 & 0.1071 & [\onlinecite{andresen}] \\
CaH$_2$ H1 & 0.3566 & 0.4274 & 0.3573 & 0.4269 & [\onlinecite{andresen}] \\
CaH$_2$ H2 & 0.9741 & 0.6773 & 0.9737 & 0.6766 & [\onlinecite{andresen}] \\
SrH$_2$ Sr & 0.2382 & 0.1109 & 0.2438 & 0.1108 & [\onlinecite{brese}]  \\
SrH$_2$ H1 & 0.3558 & 0.4278 & 0.3570 & 0.4281 & [\onlinecite{brese}]  \\
SrH$_2$ H2 & 0.9732 & 0.6787 & 0.9693 & 0.6825 & [\onlinecite{brese}] 
\end{tabular}
\label{table2}
\end{table}

\begin{table}[tbp]
\caption{LDA energies of formation on a per H$_2$ basis, assuming
full decomposition into separated elemental metals and $H_2$. (minus
means that the formation is exothermic).
$\Delta H_{static}$ denotes the LDA energy with no correction
for zero point motion, $\bar{\omega}$ is the average phonon frequency
estimated from LDA calculations (see text) and $\Delta H$(kJ mol$^{-1}$)
is the zero-point corrected formation energy for the hydride. The
LDA vibrational frequency of $H_2$ is used in this calculation.}
\begin{tabular}{lccc}
Compound & $\Delta H_{static}$(kJ mol$^{-1}$) & $\bar{\omega}$(cm$^{-1}$)
& $\Delta H$(kJ mol$^{-1}$) \\
\hline
Mg$_2$FeH$_6$ & -147 & 1089 & -133 \\
Ca$_2$FeH$_6$ & -221 & 1033 & -209 \\
Sr$_2$FeH$_6$ & -211 &  871 & -205 \\
Mg$_2$RuH$_6$ & -147 &  968 & -137 \\
Ca$_2$RuH$_6$ & -229 & 1013 & -218 \\
Sr$_2$RuH$_6$ & -221 & 1006 & -210 \\
Sr$_2$OsH$_6$ & -216 & 1023 & -205 \\
MgH$_2$       &  -90 &  954 &  -81 \\
CaH$_2$       & -219 &  767 & -217 \\
SrH$_2$       & -207 &  670 & -208 
\end{tabular}
\label{table3}
\end{table}

\begin{table}
\caption{Calculated frequences $\omega$
(cm$^{-1}$) and displacement patterns
for zone-center modes of Sr$_2$RuH$_6$, $a=14.361 a_0$ compatible with
$R32$ symmetry.
The displacements are ($\alpha_{Sr},\alpha_{Sr},\alpha_{Sr}$)
for the first Sr (in $R32$ symmetry),
($\beta_{Sr},\beta_{Sr},\beta_{Sr}$) for the second Sr, 
($\gamma_{Ru},\gamma_{Ru},\gamma_{Ru}$) for Ru,
($\delta_{H1},\epsilon_{H1},\epsilon_{H1}$) for the first 
type (in $R32$ symmetry) of H (three atoms), and
($\delta_{H2},\epsilon_{H2},\epsilon_{H2}$) for the second 
type of H (three atoms). The $\delta_{H}$ are Ru-H bond stretch coordinates,
while $\epsilon_{H}$ are Ru-H bond shears.
}    
\begin{tabular}{lccccccccc}
\multicolumn{2}{c}{$\omega$}
& \multicolumn{1}{c}{$\alpha_{\text{Sr}}$}
& \multicolumn{1}{c}{$\beta_{\text{Sr}}$}
& \multicolumn{1}{c}{$\gamma_{\text{Ru}}$}
& \multicolumn{1}{c}{$\delta_{\text{H1}}$} 
& \multicolumn{1}{c}{$\epsilon_{\text{H1}}$} 
& \multicolumn{1}{c}{$\delta_{\text{H2}}$} 
& \multicolumn{1}{c}{$\epsilon_{\text{H2}}$} \\
\hline
   \multicolumn{2}{c}{126}
  & 0.073 & -0.078  & 0.002 &  0.001 &  0.013 &  0.002 & -0.009
  \\
   \multicolumn{2}{c}{141}  
 &   -0.051 & -0.045 &  0.075  & 0.075 &  0.037 &  0.071 &  0.041
  \\
   \multicolumn{2}{c}{738}
  & 0.003 &  -0.000 & -0.005 & -0.024 & -0.673 & -0.079 & 0.195
  \\
   \multicolumn{2}{c}{777}
 & -0.002 & -0.003 & -0.011 & -0.087 &  0.188 & -0.012 &  0.670
  \\
   \multicolumn{2}{c}{1433}
 &  0.000 &  0.000 &  0.015 & -0.632 & -0.055 & -0.750  & -0.020
  \\
   \multicolumn{2}{c}{1742}
 &  0.000 &  0.000 &  0.002  & -0.757 &  0.012 &  0.643  & -0.044

\end{tabular}
\label{f-table1}
\end{table}

\begin{table}
\caption{
Phonon frequencies and displacement patterns,
as in Table~\protect\ref{f-table1} but for Pb$_2$RuH$_6$, $a=14.361 a_0$;
frequencies, $\omega$ are in cm$^{-1}$.}    
\begin{tabular}{lcccccccc}
\multicolumn{2}{c}{$\omega$}
& \multicolumn{1}{c}{$\alpha_{\text{Pb}}$}
& \multicolumn{1}{c}{$\beta_{\text{Pb}}$}
& \multicolumn{1}{c}{$\gamma_{\text{Ru}}$}
& \multicolumn{1}{c}{$\delta_{\text{H1}}$} 
& \multicolumn{1}{c}{$\epsilon_{\text{H1}}$} 
& \multicolumn{1}{c}{$\delta_{\text{H2}}$} 
& \multicolumn{1}{c}{$\epsilon_{\text{H2}}$} \\
\hline
   \multicolumn{2}{c}{23 }
& -0.023 & -0.019  & 0.089  & 0.082 &  0.038 &  0.077 &  0.045
  \\
   \multicolumn{2}{c}{52}
&  0.049 & -0.048 &  0.005 &   0.005 &  0.025 &  0.004 &  -0.017
  \\
   \multicolumn{2}{c}{552}
& -0.002 & -0.002 & -0.010 & -0.062  & 0.518 &  -0.013 &  0.469
  \\
   \multicolumn{2}{c}{632}
 & 0.002 & -0.002 & -0.001 & -0.022 & -0.473 &  0.041 &  0.520
  \\
   \multicolumn{2}{c}{1265}
 & 0.000 &  0.000  & 0.013 & -0.646 & -0.028 & -0.745 & -0.008
  \\
   \multicolumn{2}{c}{1626}
 & -0.000 &  0.000  & 0.002 & -0.749 &  0.002  & 0.654 & -0.040
 \\
\end{tabular}
\label{f-table2}
\end{table}   

\end{multicols}

\begin{references}
\bibitem{tet-note}
A tetrahedral crystal field might
also be thought to
be favorable as it is also a geometry that is favorable for
large large ligand field splittings and hybridization,
but in that case only up to 12 electrons would be accomodated
-- 8 in the 4 H $1s$ orbitals and 4 in the now lower $e_g$ manifold.
However, this is not compatible with the electron count of Fe and Mg.
\bibitem{pb-lat}
The calculated LDA lattice parameters of Pb$_2$RuH$_6$ and
Sr$_2$RuH$_6$ are 14.10 $a_0$ and 14.17 $a_0$, respectively, as
compared to the experimental lattice parameter of 14.3619 $a_0$
for Sr$_2$RuH$_6$.
\end{references}
\end{document}